\documentclass[sigconf]{acmart}

\def\BibTeX{{\rm B\kern-.05em{\sc i\kern-.025em b}\kern-.08emT\kern-.1667em\lower.7ex\hbox{E}\kern-.125emX}}
    
\copyrightyear{2019} 
\acmYear{2019} 
\acmConference[RecSys '19]{Thirteenth ACM Conference on Recommender Systems}{September 16--20, 2019}{Copenhagen, Denmark}
\acmBooktitle{Thirteenth ACM Conference on Recommender Systems (RecSys '19), September 16--20, 2019, Copenhagen, Denmark}
\acmPrice{15.00}
\acmDOI{10.1145/3298689.3347008}
\acmISBN{978-1-4503-6243-6/19/09}

\usepackage{color}
\usepackage{amsmath}
\usepackage{cases}
\usepackage{amsfonts}
\usepackage[ruled]{algorithm2e}

\makeatletter
\newcommand{\figcaption}[1]{\def\@captype{figure}\caption{#1}}
\newcommand{\tblcaption}[1]{\def\@captype{table}\caption{#1}}
\makeatother
\newcommand\figref[1]{{Figure \ref{fig:#1}}}
\newcommand\tabref[1]{{Table \ref{tab:#1}}}
\newcommand{\argmin}{\mathop{\rm arg~min}\limits}
\usepackage{subfigure}

\begin{document}

%
\title{Greedy Optimized Multileaving for Personalization}

%

\author{Kojiro Iizuka}
\affiliation{%
  \institution{Gunosy Inc.}
  \city{Tokyo}
  \country{Japan}
}
\email{kojiro.iizuka@gunosy.com}

\author{Takeshi Yoneda}
\affiliation{%
  \institution{Gunosy Inc.}
  \city{Tokyo}
  \country{Japan}
}
\email{takeshi.yoneda@gunosy.com}

\author{Yoshifumi Seki}
\affiliation{%
  \institution{Gunosy Inc.}
  \city{Tokyo}
  \country{Japan}}
\email{yoshifumi.seki@gunosy.com}

\renewcommand{\shortauthors}{K. Iizuka, et al.}

%
\begin{abstract}
Personalization plays an important role in many services. 
To evaluate personalized rankings, online evaluation, such as A/B testing, is widely used today.
Recently, {\it multileaving} has been found to be an efficient method for evaluating rankings in information retrieval fields.
This paper describes the first attempt to optimize the multileaving method for personalization settings.
We clarify the challenges of applying this method to personalized rankings. 
Then, to solve these challenges, we propose greedy optimized multileaving (GOM) with a new {\it credit feedback function}.
The empirical results showed that GOM was stable for increasing ranking lengths and the number of rankers.
We implemented GOM on our actual news recommender systems, and compared its online performance.
The results showed that GOM evaluated the personalized rankings precisely, with significantly smaller sample sizes ($<1/10$) than A/B testing.

\end{abstract}

%
%


%

%
\maketitle

\section{INTRODUCTION}
\label{introduction}
Personalization plays an important role in various web services, such as e-commerce, streaming, and news \cite{airbnb, googlenews, yahoopnews}.
A/B testing has been widely adopted to evaluate the effectiveness of personalization algorithms. 
For a precise evaluation, however, A/B testing requires a large number of users; the more the number of algorithms for A/B testing increases, the more users are needed. 

Recently, multileaving has been found to be an efficient method for evaluating rankings in information retrieval fields \cite{largeinterleaved}.
Interleaving evaluates two rankings with a small sample size, while multileaving extends the evaluation to three or more rankings.
However, studies that examine the application of multileaving for personalized rankings are limited \cite{netflixinterleaving}.

This paper describes our first attempt to optimize multileaving for personalized rankings.
The paper's contributions are as follows:
\begin{itemize}
    \item \textbf{New Problems}: Clarifies the challenges of applying the multileaving method to personalized rankings;
    \item \textbf{Algorithm}: Proposes the greedy optimized multileaving (GOM) method with the {\it personalization credit function} to solve these problems; and
    \item \textbf{Stability and Sensitivity}: Confirms the stability and sensitivity of GOM throughout offline and online experiments.
\end{itemize}
We achieved high sensitivity with low computation cost on the personalized rankings, which differed for each user and each time.
In addition, we ensured stability over the number of rankers and ranking length.   
These results show that the proposed method can evaluate many algorithms or hyperparameters efficiently in practical environments, such as Social Network Service (SNS), news, and Consumer to Consumer (CtoC) market services.

\section{RELATED WORK AND CHALLENGES}
\label{relatedwork}
\subsection{Multileaving Method}
Multileaving is a method that evaluates multiple rankings using user click feedback \cite{multileavedcomparisonscikm}.
The typical steps of a multileaving evaluation are as follows.

The first step is to input multiple ranking sets ${I} = \{I_1,I_2,\dots,I_n\}$.
These input rankings are generated from {\it rankers}.
The second step is to generate output ranking sets ${O=\{O_1,O_2,\dots,O_m\}}$.
The third step is to show each output ranking $O_k$ to users with probability $p_k$.
The fourth step is to aggregate the user click {\it credit} and evaluate which input ranking is better. 
Basically, the ranking is evaluated as better when the sum of credit is higher. 

Each multileaving method consists of three components: (1) a way to construct $O$ from $I$, (2) probability $p_k$ for each output ranking $O_k$, and (3) a credit function $\delta$.
In this paper, we represent the i-th item of ranking $I_j$ and $O_j$ as $I_{j,i}$ and $O_{j,i}$; the user click credit as $\delta( {\it O_{k,i}}, {\it I_{j}} )$; and the length of $O_k$ as $l$.
In the next section, we describe two established multileaving methods, team draft multileaving (TDM) \cite{multileavedcomparisonscikm} and optimized multileaving (OM) \cite{optimized}, in detail.

\subsection{Team Draft Multileaving}
TDM is a method that selects an input ranking randomly and adds an item to the output ranking that was not previously chosen \cite{tdmoriginal}.
This process is then repeated until the multileaved ranking is of sufficient length.
In the credit aggregation of TDM, rankers are credited for each click on an item drawn from the corresponding ranker; however, because rankers are credited only for clicks on items drawn from the corresponding ranker, TDM tends to be unstable over increasing ranker sizes \cite{improved}.

\subsection{Optimized Multileaving}

OM generates the output rankings, and solves an optimization problem formulated by {\it sensitivity} and {\it bias} \cite{optimized}. 
The output ranking's sensitivity discriminates the differences in effectiveness between input rankings; meanwhile, the bias of the output rankings measures the differences between the expected credits of the input rankings for random clicks \cite{multileavedcomparisonqa,multileavedcomparisonscikm}.
In OM, the sensitivity that constrains the bias is maximized.
By solving this optimization problem, OM achieves more sensitivity than TDM. 
In this paper, similar to \cite{multileavedcomparisonscikm, multileavedcomparisonqa}, we use insensitivity instead of sensitivity.
Insensitivity ($\sigma$) is defined by the credit sum and the mean of credit $\mu_k$.
$
\sigma_k^2 = \sum_{j=1}^{n}\big( \sum_{i=1}^{l} f(i)\delta( {\it O_{k,i}}, {\it I_{j} } \big) - \mu_{k} \big)^2,
$
where $f(i)$ is the probability with which a user clicks on the i-th item, and the mean of credit $\mu_k = \frac{1}{n} \sum_{j=1}^n\sum_{i=1}^l f(i)\delta( {\it O_{k,i}}, {\it I_{j} } \big) $.
The formulation of OM is as follows \cite{multileavedcomparisonqa}:
\begin{align}
 \label{eq:omproblem} 
&\min_{p_k} \qquad \alpha \sum_{r=1}^{l} \lambda_{r} + \sum_{k=1}^m p_{k} \sigma_k^2& \nonumber \\
&\text{subject to}& \notag\\
&(\forall r, j,j') | \sum_{k=1}^{m}p_k\sum_{i=1}^{r}\delta( {\it O_{k,i}}, {\it I_{j}} ) - \sum_{k=1}^{m}p_k\sum_{i=1}^{r}\delta( {\it O_{k,i}}, {\it I_{j'}} ) | \leq \lambda_r & \notag \\
&\qquad \qquad \sum_{k=1}^m p_k = 1, \quad  0 \leq p_k \leq 1 \quad(k=1,\dots,m). & \notag\\
\end{align}

We follow the original work \cite{multileavedcomparisonscikm}, and use $f(i) = 1/i$.
The bias term is added to the objective function, and is adjusted with hyperparameter $\alpha$.
$\alpha$ is the only hyperparameter in this formulation.
$\lambda_r$ is the maximum difference of the expected credits in any input rankings pairs, and represents the bias.
If $\lambda_r$ is close to zero, the bias becomes small.

In previous studies, the click credit function was defined by the input ranking's position \cite{optimized, multileavedcomparisonscikm, multileavedcomparisonqa}.
For example,
\begin{eqnarray}
\delta( {\it O_{k,i}}, {\it I_{j}}) = \dfrac{1}{rank({\it O_{k,i}}, {\it I_{j}})},
\label{eq:inversecredit} 
\end{eqnarray}
where $rank({\it O_{k,i}, \it I_{j}})$ represents the position of item ${\it O_{k,i}}$ in ranking ${\it I_{j}}$.
If there is no item ${\it O_{k,i}}$  in ranking ${\it I_{j}}$, then the credit value is $1/(|{\it I_{j}}|+1)$.
We call this credit function {\it inverse credit}.
By definition, the deeper the click position, the weaker the credit.
Another well-known definition is  $\delta( {\it O_{k,i}}, {\it I_{j}}) = -rank({\it O_{k,i}},  {\it I_{j}}) $ if $O_{k,i}  \in I_{j} $; otherwise, $-(|I_j|+1)$. 
In this case, the credit difference between the top and bottom items is too big, compared to (top item, top item) or (bottom item, bottom item).
This difference induces noise for the credit evaluation. 
The empirical results, described in Section \ref{experiments}, indicated that when credit above is used, OM tended to become unstable as the ranking length increased.

\subsection{Challenges}
The challenges of applying the multileaving method to personalization are as follows:
\begin{enumerate}
    \renewcommand{\labelenumi}{C\arabic{enumi}:}
    \item Achieving high sensitivity with low computation cost for the rankings, which differed for each user and each time; and
    \item Ensuring stability over the number of rankers and the ranking length.
\end{enumerate}

OM maximizes sensitivity by assuming that the same rankings are inputted multiple times.
In personalization, however, input rankings differ for each user. 
Furthermore, input rankings vary according to time stamps in environments where new items rapidly increase, such as SNS, news, and CtoC market services.
Therefore, the ranking differs for each user and each time.
In these cases, OM cannot be pre-calculated, and requires low computation cost to show multileaved output rankings to users in real-time.

Unstability over the number of rankers is known to occur in TDM \cite{improved, multileavedcomparisonscikm}.
This is a serious problem for personalization, because there are usually many algorithm candidates and hyperparameters.
In addition, the empirical results showed that OM was unstable over the ranking length when ordinary click credit is used.
This is critical because many services have long ranking length to show as many personalized items as possible.
Thus, for personalization, we should ensure stability over the number of rankers and the ranking length.

\section{GREEDY OPTIMIZED MULTILEAVING}
\label{proposal}

\subsection{Formulation}
To solve the first challenge, we propose a new formulation using the characteristic of personalized ranking.
Personalized ranking is shown to the user a few times, because it differs for each user and each time in the situation when the number of new items increase rapidly.
In other words, the problem in OM for personalization is to select a output ranking $\it{O_k}$ from overall output ranking $\it{O}$.
Therefore, we can consider ranking output probability as a one-hot vector, $p_k=1$, if $\it{O_k}$ is selected; otherwise, $p_{-k}=0$.
Equation \eqref{eq:omproblem} can be simplified as follows:
\begin{align}
&\argmin_{k} \qquad \alpha \sum_{r=1}^{l} \lambda_{r} + \sigma_k^2& \notag \\
&\text{subject to}&  \notag \\
&\qquad (\forall r, j,j') | \sum_{i=1}^{r}\delta( {\it O_{k,i}}, {\it I_{j}} ) - \sum_{i=1}^{r}\delta( {\it O_{k,i}}, {\it I_{j'}} ) | \leq \lambda_r. & \notag
\end{align}To solve this problem, we can use the greedy strategy. In this paper, we call this formulation greedy optimized multileaving (GOM).
GOM has a low computation cost, because it does not need to solve a linear problem.
This strategy is fast enough to compute in real-time in production recommender systems.
Also, our implementation is publicly available \footnote{https://github.com/mathetake/intergo}.

\subsection{Assigning Credit}
To solve the second challenge, we propose a new definition for the credit function:\begin{eqnarray}
\delta( {\it O_{k,i}}, {\it I_{j}}) = -|\{j^{'}  |  rank({\it O_{k,i}},  {\it I_{j^{'}}})  \le rank({\it O_{k,i}},  {\it I_{j}}) \}|,
\label{eq:newcredit} 
\end{eqnarray}If there is no item ${\it O_{k,i}}$  in ranking ${\it I_{j}}$, then the credit value is $-(|{\it I_{j}}|+1)$.
We call this credit personalization credit.

This definition is interpreted as considering a mutual interaction with input rankings, and gives credit to multiple rankings per click.
Because relative ranking orders are used instead of the rankings' absolute position, the credits are calculated without position noise.
For example, we set $I_1$=[1,2,$\dots$,99,100,101,102], $I_2$=[1,2,$\dots$,99,101,102,100], $I_3$=[1,2,$\dots$,99,102,100,101], and $O_k$=[1,2,$\dots$,99,102,101,100].
When $101$ is clicked, the personalization credit values are $\delta(101, I_1) = -2$, $\delta(101, I_2) = -1$, and $\delta(101, I_3) = -3$.
Conversely, the inverse credits are $\delta(101, I_1) = 1/101 = 0.0099$, $\delta(101, I_2) = 1/100 = 0.01$, and $\delta(101, I_3) = 1/102 = 0.0098$.
Each absolute inverse credit value is much smaller and closer than personalization credits.

\section{EXPERIMENTS}
\label{experiments} 
\begin{table*}[tp]
  \caption{Differences between the sum of credits. The values of GOM-P and TDM were all positive, while some values of GOM-I (written in blue) were negative. The GOM-P and TDM's results were consistent with previous CTR results of A/B testing (algo-A $<$ algo-B $<$ algo-C $<$ algo-D $<$ algo-E), but  GOM-I was inconsistent. This means that the inverse credit function which was often used in previous studies is not appropriate for long ranking.}
  \label{tab:table3}
  \begin{tabular}{|l||rrrr|rrrr|rrrr|} \hline
     &  \multicolumn{4}{c|}{GOM-I} & \multicolumn{4}{c|}{GOM-P} & \multicolumn{4}{c|}{TDM} \\ 
     & algo-A & algo-B & algo-C  & algo-D & algo-A & algo-B & algo-C & algo-D & algo-A & algo-B & algo-C  & algo-D \\ \hline \hline
    algo-B & \textcolor{blue}{-14,962} & &  &  & 21,111 &  &  &  & 615 & &  &     \\ \hline
    algo-C & \textcolor{blue}{-21,642} & \textcolor{blue}{-6,680} &  &  & 36,561 & 15,450 &  &   & 1,259 & 644 &  &     \\ \hline
    algo-D & \textcolor{blue}{-24,179} & \textcolor{blue}{-9,217} & \textcolor{blue}{-2,537}  & & 44,484 & 23,373 & 7,923  &   & 1,812 & 1,197 & 553  &      \\ \hline
    algo-E & \textcolor{blue}{-5,246} & 9,716 & 16,396 & 18,933 & 52,597 & 31,486 & 16,036  & 8,113 &     2,117 & 1,502 & 858 & 305   \\ \hline
  \end{tabular}
\end{table*}

\subsection{Offline Experiment Settings}
In the offline experiment, we simulated user clicks, and evaluated several methods, which are compared below:
\begin{itemize}
    \item TDM: Described in Section \ref{relatedwork};
    \item GOM-I: GOM, using the inverse credit \eqref{eq:inversecredit}; and
    \item GOM-P: GOM, using the personalization credit \eqref{eq:newcredit}.
\end{itemize}
These experiments assumed a practical environment that requires a low computation cost to generate rankings in real-time; therefore, we did not use OM for the performance comparisons.
We used TDM for the performance comparison described in Section \ref{experiments} because TDM has been examined in online settings\cite{multileavedcomparisonqa}.

{
\begin{algorithm}
\caption{User click simulation in personalized setting}
\label{simulation}
\SetKwInOut{Input}{input}
\SetKw{Return}{return}

\Input{the number of rankers $n$, ranking length $l$}
$win=0$ \\
\For{$i=1,\dots, numeval$}{
	Select ranking index $r$ randomly from $1$ to $n$ \\
    \lFor{$k=1,\dots, n$}{$credit[k]=0$}
	\For{$j=1,\dots, numclick$}{
    	$InitialRanking$ = generateRankingRandomly($l$) \\
		\lFor{$k=1,\dots, n$}{$\it I_k$ = Shuffle($InitialRanking$)} 
       Get MultileavedRanking $\it O$ from $\it I$ \\
       Select one $item$ from $I_r$ at the top $x\%$ position randomly\\
       Click $item$ in ${\it O}$ and update sum of $credit$ for all $\it I$
	}
    $win$ += |\{k|$credit[k] > credit[r]$ \} | 
} 
$accuracy$ = $win/(numeval*(n-1))$ \\
\Return $accuracy$; 
\end{algorithm}
}

The simulation steps are shown in Algorithm \ref{simulation}.
We fixed the constant values $numeval=100$, $numclick=100$, number of output rankings = $10$, and click bias probability $x=80\%$.
We evaluated the accuracy over the number of rankers ${2,3,...20}$ when the ranking length was fixed at 10.
We also evaluated the accuracy of ranker lengths ${5, 15, ..., 195}$ when the number of rankers was fixed at 3.

Next, we evaluated insensitivity and bias.
Insensitivity $\sigma_k$ was divided by the square of the average credit $\mu_k^2$, because the insensitivity is proportional to it.

\subsection{Offline Experiment Results and Discussion}

\begin{figure}[tp]
\centering
	\begin{tabular}{cc}
	\begin{minipage}[t]{.20\textwidth}
\includegraphics[width=40mm,height=30mm]{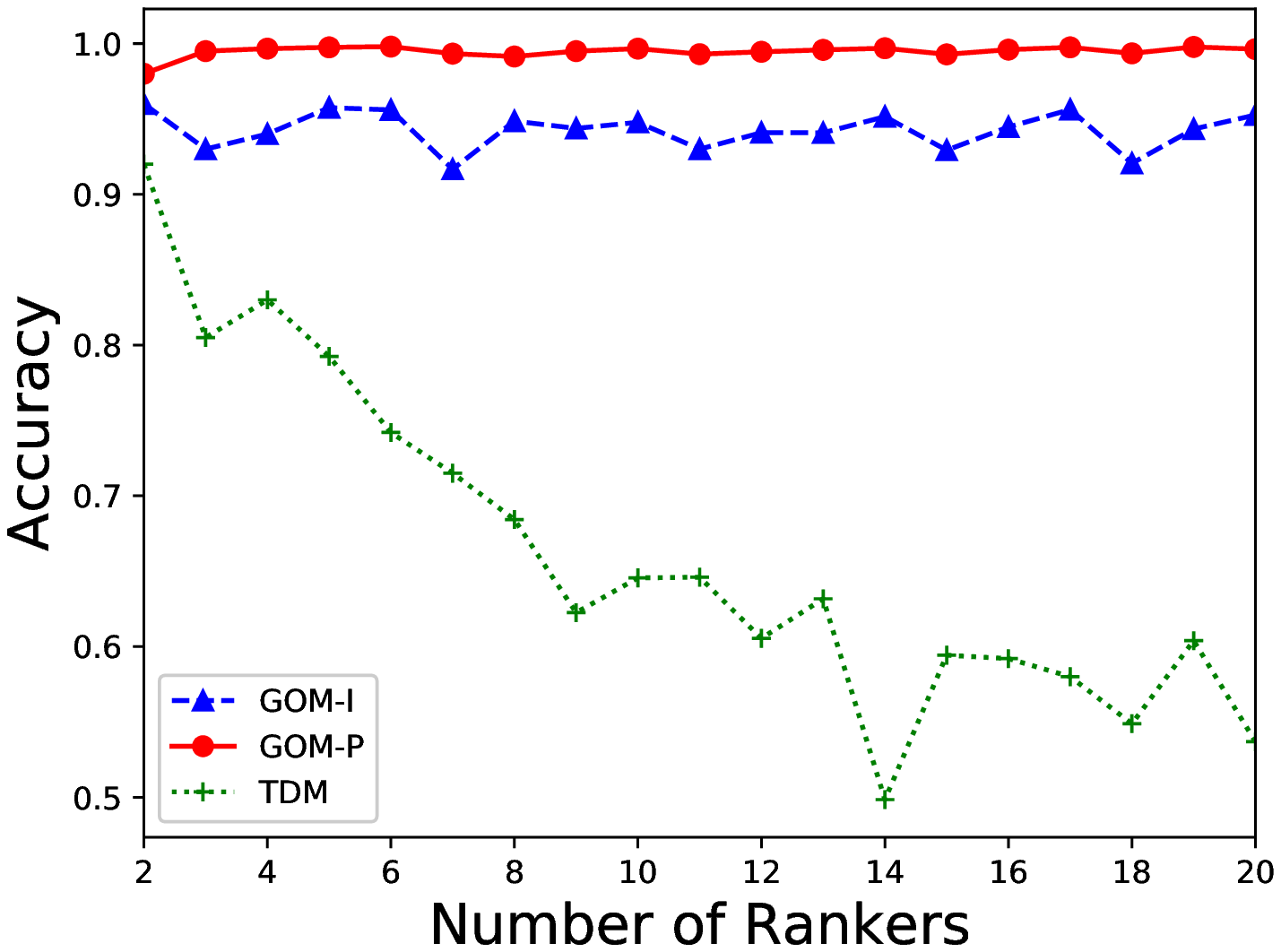}
\caption{Accuracy versus the number of rankers for the fixed ranking length fixed at 10 using the random click simulation (averaged over 100 runs).}
\label{fig:rankersize}
\end{minipage}
	&
	\begin{minipage}[t]{.20\textwidth}
\includegraphics[width=40mm,height=30mm]{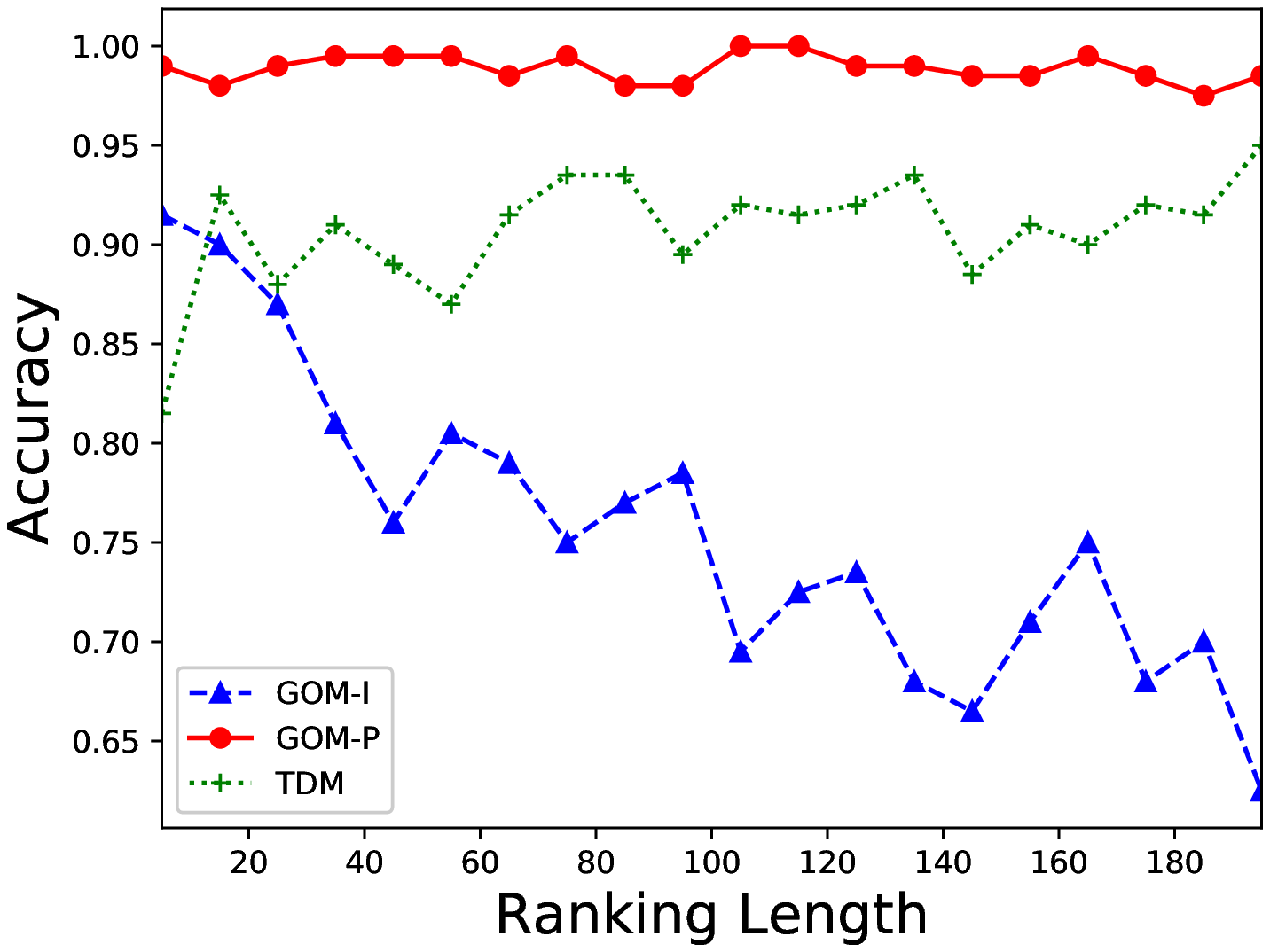}
\caption{Accuracy versus the ranking lengths for the fixed number of rankers fixed at 3 using the random click simulation (averaged over 100 runs). }
\label{fig:rankinglen}
\end{minipage}
	\end{tabular}
\end{figure}

\begin{figure}[tp]
\centering
\label{fig:pvalue}
	\begin{tabular}{cc}
	\begin{minipage}[t]{.20\textwidth}
\includegraphics[width=40mm,height=30mm]{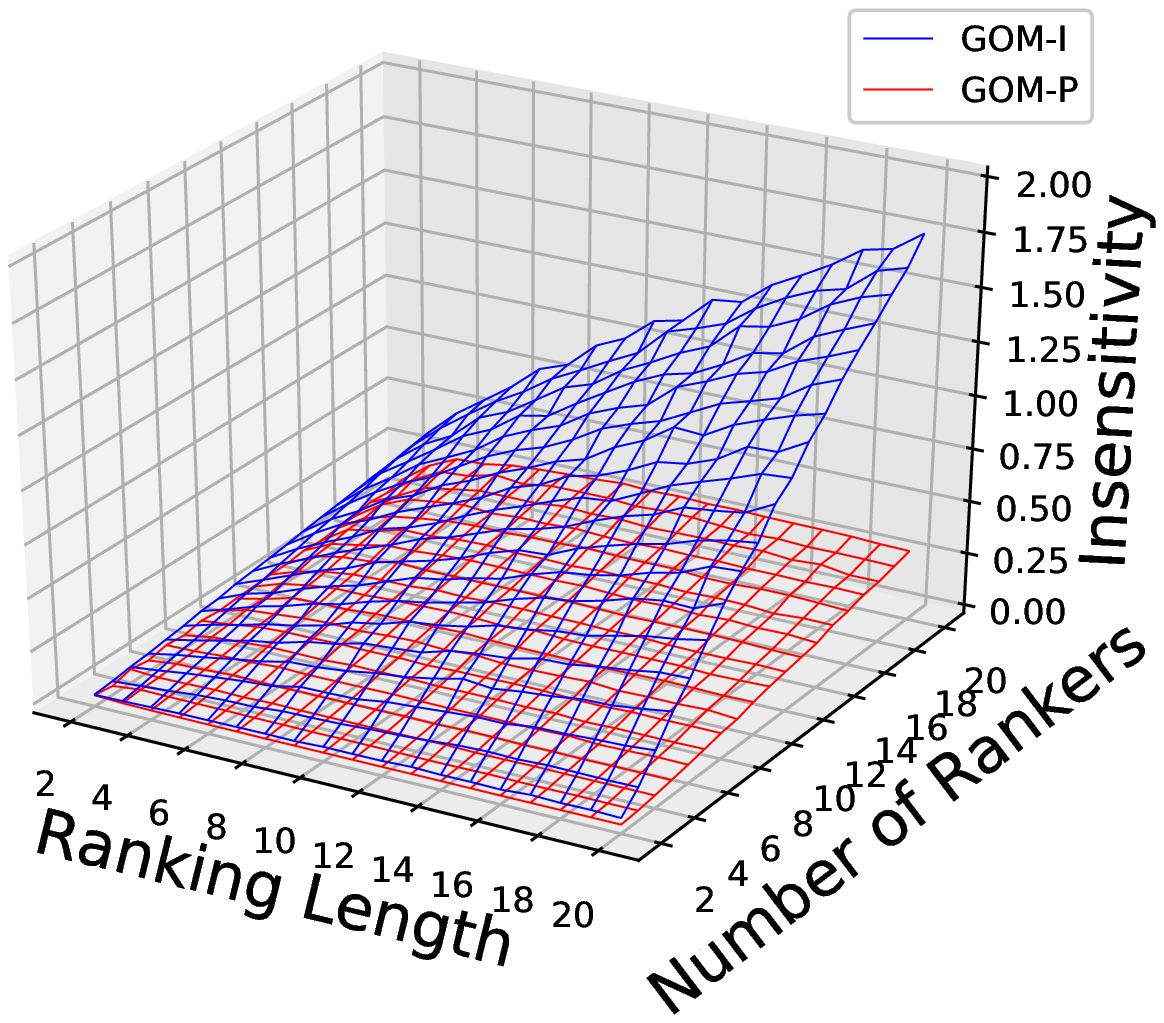}
\caption{Insensitivity versus the number of rankers and the ranking lengths  (averaged over 100 runs) .}
\label{fig:insensitivity_rankersize_length}
\end{minipage}
&
\begin{minipage}[t]{.20\textwidth}
\includegraphics[width=40mm,height=30mm]{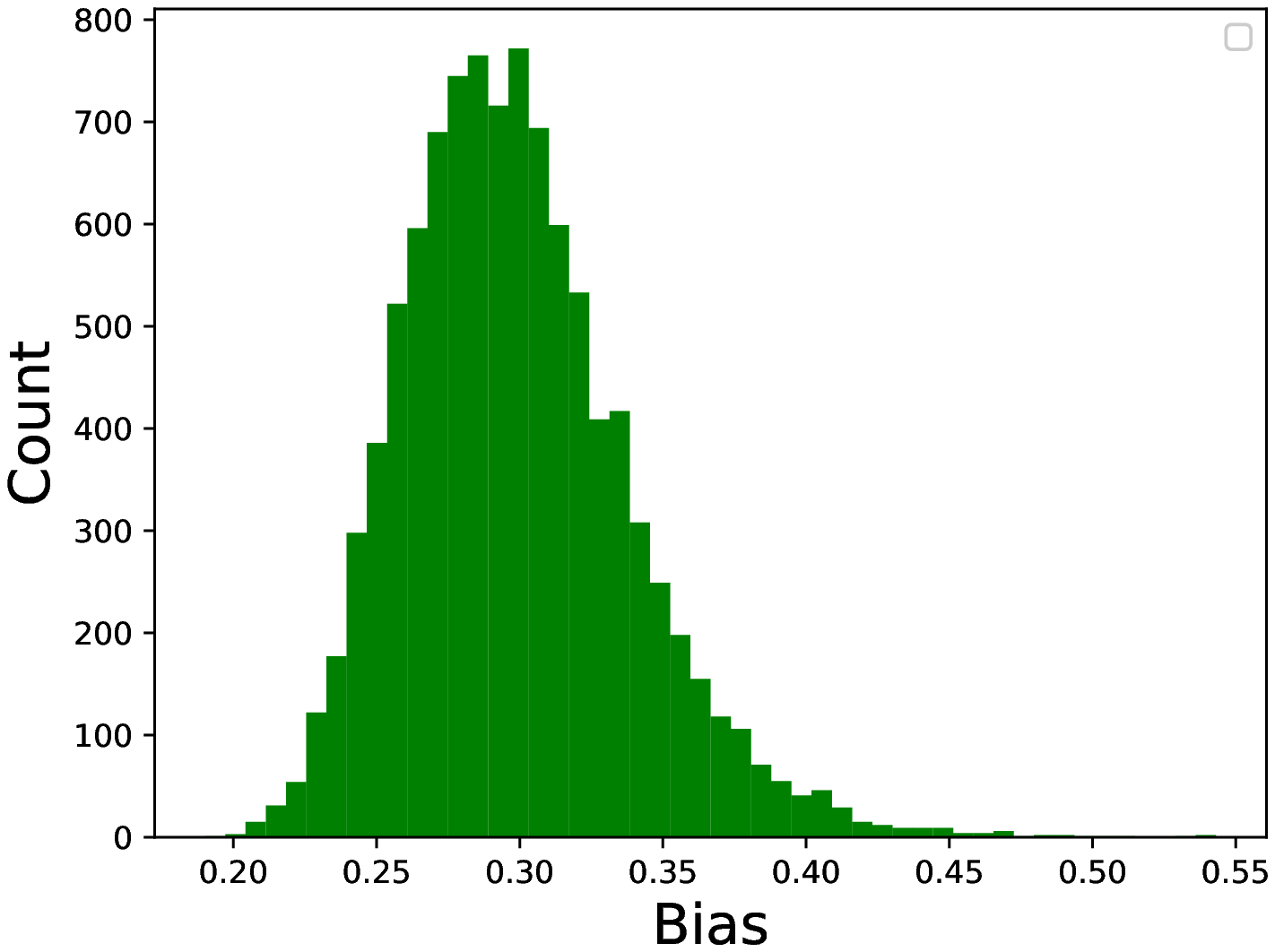}
\caption{Bias distribution on generating GOM-P rankings 10,000 times.} 
\label{fig:biascount}
\end{minipage}
\end{tabular}
\end{figure}

\figref{rankersize} shows that GOM-P and GOM-I were more accurate compared to TDM.
When the number of rankers increased, TDM's accuracy decreased. 
TDM credit caused this inaccuracy.

\figref{rankinglen} shows that the GOM-P and TDM methods had higher accuracy compared to GOM-I.
When the ranking length increased, GOM-I's accuracy decreased. The noise of the inverse credit at the ranking's deep position of the ranking induced this inaccuracy.
In contrast, TDM and GOM-P were stable over the ranking lengths.

\figref{insensitivity_rankersize_length} shows the insensitivity for the number of rankers and the ranking length.
GOM-P was sensitive compared to GOM-I in these cases; therefore, personalization credit achieved high sensitivity over the number of rankers and ranking lengths.
This sensitivity resulted in the higher accuracy of GOM-P.

\figref{biascount} shows the bias distribution of GOM-P, which appears to be a normal distribution.
The ideal bias distribution is that all biases are the same value at some point.
The standard deviation of GOM-P was 0.039422, and the mean was 0.298570.

Interestingly, we found that hyperparameter $\alpha$ did not affect the accuracy, insensitivity, or standard deviation of bias.
This means that we can set $\alpha = 0$.
Then, to get a multileaved ranking, we only have to minimize the insensitivity in GOM; no parameter tuning is needed.

\subsection{Online Experiment Settings}
We conducted an online experiment on Gunosy,\footnote{https://gunosy.com} one of the most popular news applications in Japan.
On this service, we recommend personalized news articles using a user click log \cite{gunosylogic}.
The topics of the news articles are broad, ranging from entertainment to political opinions.

We prepared five personalization algorithms, and evaluated the effectiveness for each multileaving method.
According to previous A/B testing, we knew the effectiveness of the algorithms as algo-A $<$ algo-B $<$ algo-C $<$ algo-D $<$ algo-E.
In this A/B testing, we used the click-through rate  (CTR) as the effectiveness of the personalization algorithms.
CTR is a metric that divides article clicks by article impressions.

The online experiment was carried out for one week.
We presented multileaved rankings for a particular portion of user requests, and received 10,826,923 article impressions.
The deepest ranking length was up to 120.

\subsection{Online Experiment Results and Discussion}
\tabref{table3} shows the credit differences between the evaluated algorithms in GOM-P, GOM-I, and TDM.
The value $21,111$ for GOM-P indicates the credit difference between algo-B and algo-A.
The differences of sum credit for GOM-P and TDM were all positive values, which means that the GOM-P and TDM's results were consistent with previous CTR results.
In contrast, some values of GOM-I written in blue were negative, and inconsistent with previous outcomes.
In this online experiment setting, the number of rankers was five, and the ranking length was longer than 100.
According to the offline experiments, this small number of rankers caused the consistent TDM results, while the long ranking length caused the inconsistent GOM-I results.

\begin{figure}[ht]
\centering
\includegraphics[width=60mm,height=45mm]{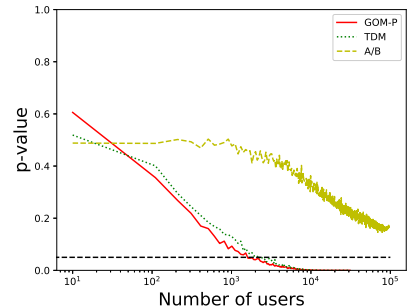}
\caption{P-values versus the number of users. The multileaving methods converged much faster than A/B testing (< 1/10).} 
\label{fig:avepvalue}
\end{figure}

\figref{avepvalue} shows that the multileaving methods converged much faster than A/B testing ($< 1/10$).
These p-values were calculated using bootstrap subsampling (sampling size = 50).
The x-axis represents user number $N$, and the y-axis represents the p-value averaged over all pairs of the algorithms.
The multileaving p-value was from paired t-tests.
In the A/B testing,  $N/2$ users were assigned per algorithm, and the p-value came from unpaired t-tests.

One of the reasons for the slow convergence of A/B testing was group bias.
The users assigned to algo-C in the A/B testing tended to be inactive. 
The multileaving method did not suffer from this group bias, because different algorithms were examined in the same user groups.

\section{CONCLUSION}
\label{conclusion}
In this paper, we proposed a new multileaving method for personalized rankings that requires a high degree of freshness, many hyperparameters, and a long ranking length for a rich user experience. 
To the best of our knowledge, few theoretical multileaving studies have been conducted for these personalized settings. 

We clarified the challenges of applying this method for personalized rankings.
We then formalized this personalized multileaving problem, and introduced a new credit feedback function.
We also conducted an offline experiment, which showed that the proposed method was stable, regardless of the number of rankers or ranking length.
Finally, we conducted an online experiment in the large-scale recommender system. 
We confirmed that the proposed method could evaluate rankings precisely using a significantly smaller sample size than A/B testing.

In future work, we will conduct further offline click simulations for various types of click bias.
In addition, we will try to apply GOM's feedback schema to speed up online learning for ranking.
We will also investigate whether the GOM results agree with specific measures, such as Normalized Discounted Cumulative Gain (NDCG).

\renewcommand{\thefootnote}{\fnsymbol{footnote}}
\footnote[0]{We would like to thank the Gunosy Product Team, Machine Learning Team, and Data Management Platform Team for their contributions to this project.}
\renewcommand{\thefootnote}{\arabic{footnote}}

\bibliographystyle{ACM-Reference-Format}
\bibliography{sample-base}
\end{document}